\documentclass[journal,twocolumn]{IEEEtran}
\usepackage{epsfig,setspace,amsmath,epsf,amssymb,bm,theorem,cite,graphicx, epstopdf,algorithm,algpseudocode,float,color,mathtools}
\usepackage[table,xcdraw]{xcolor}
\usepackage{multirow}
\usepackage{caption}
\usepackage{subcaption}
\usepackage{comment}

\IEEEoverridecommandlockouts
\allowdisplaybreaks

\begin{document}

\title{Private Information Retrieval and Its Applications: An Introduction, Open Problems, Future Directions}

\author{Sajani Vithana \qquad Zhusheng Wang \qquad Sennur Ulukus\\
	\normalsize Department of Electrical and Computer Engineering\\
	\normalsize University of Maryland, College Park, MD 20742\\
	\normalsize \emph{spallego@umd.edu} \qquad \emph{zhusheng@umd.edu} \qquad \emph{ulukus@umd.edu}}

\maketitle

\begin{abstract}
Private information retrieval (PIR) is a privacy setting that allows a user to download a required message from a set of messages stored in a system of databases without revealing the index of the required message to the databases. PIR was introduced under computational privacy guarantees, and is recently re-formulated to provide information-theoretic guarantees, resulting in \emph{information theoretic privacy}. Subsequently, many important variants of the basic PIR problem have been studied focusing on fundamental performance limits as well as achievable schemes. More recently, a variety of conceptual extensions of PIR have been introduced, such as, private set intersection (PSI), private set union (PSU), and private read-update-write (PRUW). Some of these extensions are mainly intended to solve the privacy issues that arise in distributed learning applications due to the extensive dependency of machine learning on users' private data. In this article, we first provide an introduction to basic PIR with examples, followed by a brief description of its immediate variants. We then provide a detailed discussion on the conceptual extensions of PIR, along with potential research directions.  
\end{abstract}

\section{Introduction}

Private information retrieval (PIR) describes an elemental privacy setting where a user downloads a single message out of a set of messages stored in multiple non-colluding and replicated databases, without revealing the identity of the downloaded message. PIR finds applications in a multitude of fields, such as, medicine, finance and national defense, to access useful data without leaking any information about the retriever's needs, intents or interests. For example, an investor may wish to download certain relevant stock market records without revealing their identities, from which information about potential investments can be leaked. Similarly, an inventor may wish to search for inventions in a patent database without revealing what is being searched for, to avoid leaking any information on their own invention prior to publication. 

PIR problem was first introduced in the seminal paper \cite{original} which provided PIR schemes and computational guarantees. While being an active area of research in computer science for many years, PIR recently has attracted significant interest in information theory with the leading paper \cite{PIR} which characterizes the capacity of PIR. The subsequent papers have characterized the capacities of various PIR settings in different scenarios. These capacity results provide fundamental limits on the performance of PIR, analogous to Shannon's capacity theorem for communication channels.

\begin{figure}[t]
    \centering
    \includegraphics[width=0.8\linewidth]{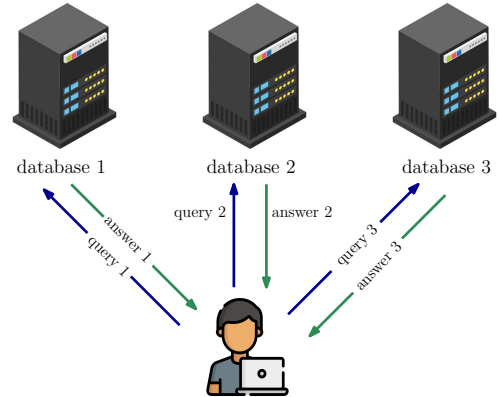}
    \caption{Communications in the PIR process.}
    \label{qanda}
    \vspace*{-0.5cm}
\end{figure}

The basic PIR setting considers a system of $N$ non-colluding databases, each storing $K$ independent messages. A user sends queries to the system of databases, with the goal of \emph{privately} requesting a desired message. Each database then sends the corresponding answer back to the user, as shown in Fig.~\ref{qanda}. The queries sent by the user must not reveal any information about the required message index to any of the databases. Formally, from the perspective of each individual database, the posterior probability that the user-required message index is $\theta$, conditioned on the query transmitted by the user, must be equal to the corresponding posterior probability with the user-required message index being $\theta^\prime$, for all $\theta^\prime \neq \theta$. This is known as the user privacy constraint. The user should be able to correctly retrieve the required message using the collection of answers from all the databases. This is known as the correctness constraint (also known as the decodability or reliability constraint), which formally states that there should be no uncertainty in the message retrieved. 

\begin{figure}[t]
    \centering  \includegraphics[width=0.82\linewidth]{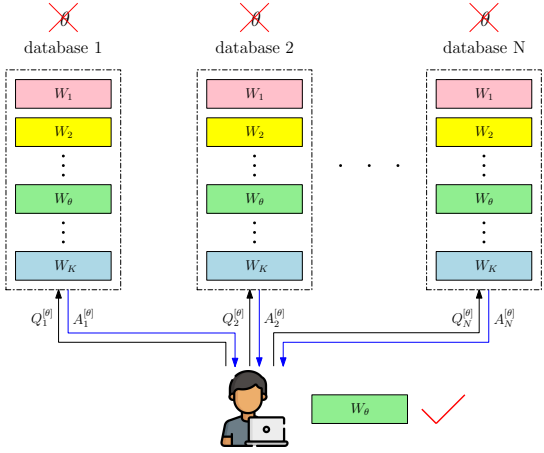}
    \caption{The system model of PIR.}
    \label{PIR_System_Model}
    \vspace*{-0.5cm}
\end{figure}

The goal of the PIR problem is to design schemes that satisfy the privacy and correctness constraints while achieving the minimum possible download cost, equivalently, the largest possible PIR rate. The download cost of a PIR scheme is defined as the total number of bits downloaded by the user from all the databases, normalized by the message size. The PIR rate is defined as the reciprocal of the PIR download cost. The system model for PIR is shown in Fig~\ref{PIR_System_Model}, where a user wants to download the message $W_{\theta}$, without revealing the message index $\theta$ to any of the databases.

\section{PIR Schemes}\label{pirschemes}

\begin{figure}[t]
    \centering
    \includegraphics[width=0.8\linewidth]{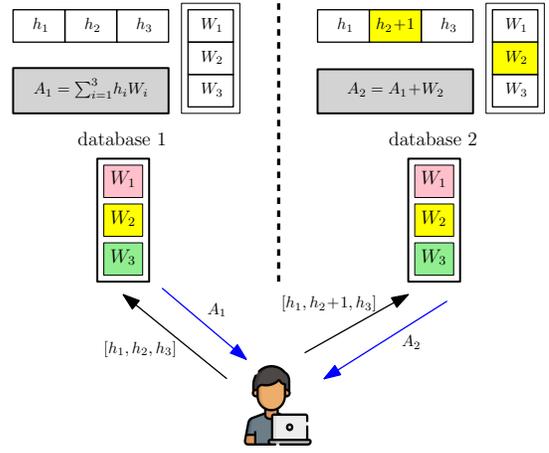}
    \caption{PIR scheme in \cite{original} for $N=2$ and $K=3$.}
    \label{eg0}
    \vspace*{-0.5cm}
\end{figure}

The first known PIR scheme that achieves information theoretic privacy is presented in \cite{original}. This scheme is based on the concept of using a pair of databases to retrieve a single symbol of the required message as illustrated in Fig.~\ref{eg0}. In the example in Fig.~\ref{eg0}, the three single-symbol messages denoted by $W_1$, $W_2$, $W_3$ are stored across two databases. These message symbols take values from a finite field $\mathbb{F}_q$. Assume that the user wants to download the second message $W_2$. As queries, the user sends $K=3$ randomly chosen symbols from $\mathbb{F}_q$ to database 1, denoted by $[h_1,h_2,h_3]$. At the same time, the user sends $[h_1,h_2+1,h_3]$ to database 2, where ``$+$'' is addition within the finite field $\mathbb{F}_q$; for instance, within the binary field, it is the XOR operation. Upon receiving the query, each database simply computes the dot product of the query and the three message symbols, and sends the result back to the user as a single symbol, as shown in Fig.~\ref{eg0}. Then, the user obtains the required message as $W_2=A_2-A_1$. User privacy is guaranteed since the databases are non-colluding, and each database simply receives a set of random symbols from $\mathbb{F}_q$. The rate of this scheme is  $R=\frac{1}{2}$, as the user downloads two symbols for one privately received symbol. 

Based on the concept introduced in \cite{original}, a more efficient PIR scheme that is compatible with arbitrary number of databases and arbitrary message lengths was proposed in \cite{ori2}. This scheme basically improves the scheme in \cite{original} by utilizing the same piece of side information (i.e., $A_1$ in Fig.~\ref{eg0}) multiple times throughout the process, as opposed to only using it once in \cite{original}. As an illustration, consider the following example with $N=3$ databases storing $K=2$ messages. The scheme is explained on a 2-symbol segment of each message, which is called a \emph{subpacket}, and is applied on all such subpackets repeatedly in an identical manner. Let $W_1=(W_{1,1},W_{1,2})$ and $W_2=(W_{2,1},W_{2,2})$ be individual subpackets of the first and second messages, respectively, each consisting of two symbols from $\mathbb{F}_q$. Let $h_{1,1}$, $h_{1,2}$, $h_{2,1}$, $h_{2,2}$ be four randomly and independently selected symbols from $\mathbb{F}_q$. Assume that the user wants to download the first message $W_1$. 

\begin{figure}[t]
    \centering
    \includegraphics[scale=0.46]{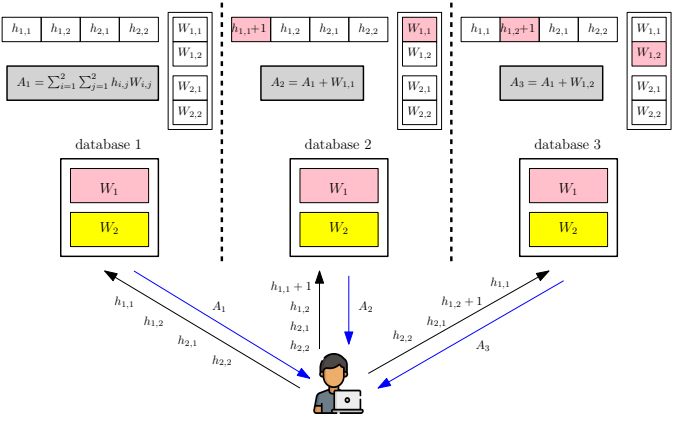}
    \caption{PIR scheme in \cite{ori2} for $N=3$ and $K=2$.}
    \label{eg1}
     \vspace*{-0.5cm}
\end{figure}

The PIR scheme for this example is shown in Fig~\ref{eg1}. The user sends the queries $Q_1=[h_{1,1},h_{1,2},h_{2,1},h_{2,2}]$, $Q_2=[h_{1,1}\!+\!1,h_{1,2},h_{2,1},h_{2,2}]$ and $Q_3=[h_{1,1},h_{1,2}\!+\!1,h_{2,1},h_{2,2}]$ to databases 1, 2 and 3, respectively. Since the three databases are non-colluding, the user privacy follows from \cite{original}. Each database computes the dot product of the received query and the four stored message symbols ($W_{1,1},W_{1,2},W_{2,1},W_{2,2}$), and sends it back to the user as the answer. The explicit expressions for the answers are shown in Fig.~\ref{eg1}. The user obtains the two symbols of the user-required message via,
\begin{align}
    W_{1,1}=A_2-A_1, \qquad
    W_{1,2}=A_3-A_1.
\end{align}
The rate achieved for this example is $R=\frac{2}{3}$, as two symbols are privately obtained by three downloads. 

For general $N$ and $K$, with this scheme, we can privately obtain $N-1$ symbols by a total of $N$ downloads. Thus, the rate of this scheme is,
\begin{align}\label{spir}
    R=\frac{N-1}{N}=1-\frac{1}{N},
\end{align}
which leads to the following question: Is this the best achievable rate or can we do better than $1-\frac{1}{N}$? 

This question is answered in \cite{PIR}, which first shows that the rate any valid PIR scheme for the general setting of $N$ databases storing $K$ messages is upper bounded by,
\begin{align}\label{pir}
    R\leq\left(1+\frac{1}{N}+\dotsc+\frac{1}{N^{K-1}}\right)^{-1},
\end{align}
which is derived using fundamental bounds in information theory. Meanwhile, there does exist optimal PIR schemes that achieve the upper bound in \eqref{pir} for any $N$ and $K$, characterizing the capacity of PIR as,
\begin{align}\label{cap}
    C_{PIR}=\left(1+\frac{1}{N}+\dotsc+\frac{1}{N^{K-1}}\right)^{-1}.
\end{align}
Note that the PIR capacity $C_{PIR}$ in (\ref{cap}) is strictly greater than the achievable rate $R$ in (\ref{spir}). 

To date, there are two primary information theoretic approaches towards achieving the capacity of PIR. The first is a \emph{deterministic} approach, inspired by the idea of blind interference alignment \cite{PIR}. The second is a  \emph{probabilistic} approach \cite{leaky,ChaoTian}, based on the idea that from the viewpoint of each database, each potential query is designed such that it could be used to retrieve any message in the message set with equal probability. These two types of schemes are described next.

The basic idea of the deterministic scheme in \cite{PIR} is to enforce message symmetry within the queries, to prevent the databases from identifying the index of the user-desired message. At the same time, the queries should be carefully designed in such a way that the user is able to decode the desired message by exploiting the unwanted message bits (side information) downloaded from different databases. 

This concept is illustrated in the following example with $N=2$ databases storing $K=2$ messages. This scheme requires the messages to be divided into \emph{subpackets} of size $N^K=4$. The subpackets corresponding to the first and second messages are denoted by $(a_1,a_2,a_3,a_4)$ and $(b_1,b_2,b_3,b_4)$, respectively. In order to retrieve the desired message $W_1$ or $W_2$, the user sends the corresponding queries to both databases with the aim of downloading the symbols as shown in Table~\ref{DPIR_N2K2}. 

\begin{table}[h]
\begin{center}
\begin{tabular}{|c|c||c|c|}
\hline
\multicolumn{2}{|c||}{Retrieve $W_1$} & \multicolumn{2}{c|}{Retrieve $W_2$} \\
\hline
DB $1$ & DB $2$ & DB $1$ & DB $2$\\ 
\hline
$a_1$ & $a_2$ & $a_1$ & $a_2$\\
$b_1$ & $b_2$ & $b_1$ & $b_2$\\
\hline
$a_3+b_2$ & $a_4+b_1$ & $a_2+b_3$ & $a_1+b_4$\\
\hline
\end{tabular}
\end{center}
\vspace{-0.5em}
\caption{Deterministic scheme in \cite{PIR} for $N=2$, $K=2$.}
\label{DPIR_N2K2}
\vspace*{-0.2cm}
\end{table}

In Table~\ref{DPIR_N2K2}, to retrieve any message, the user first downloads a single symbol of each of the two messages from both databases. Then, the user downloads sums in the form of $a+b$ from both databases to satisfy the message symmetry. If the desired message is $W_1$, $a$ is a new symbol of $W_1$ and $b$ is an already downloaded symbol from the other database and vice versa. Each unwanted download is used as side information in another database to increase the efficiency of the PIR process. User privacy is guaranteed by maintaining message symmetry among all types of queries, i.e., each individual database always receives queries requesting a single bit of each of the two messages, and a sum of two new bits of the two messages, irrespective of the user's message requirement.\footnote{The subscripts of $a$ and $b$ are permuted by the user, prior to sending the queries to ensure that no information is leaked by them.} The rate achieved in this example for any given message requirement is $R=\frac{2}{3}$, since $6$ bits are downloaded in total to obtain $4$ bits of the required message. This rate equals the capacity in \eqref{cap} when $N=2$ and $K=2$. 

\begin{figure}[t]
    \centering
    \includegraphics[width=0.8\linewidth]{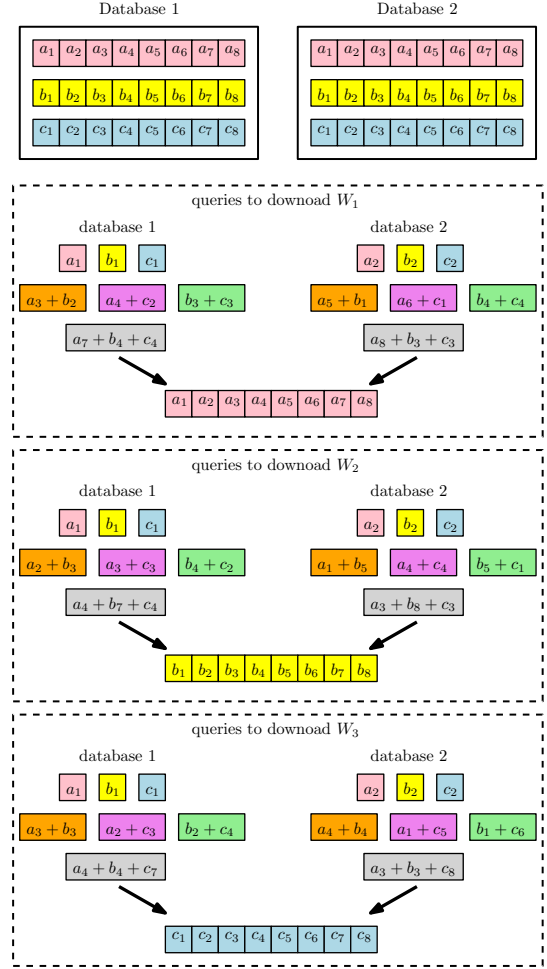}
    \caption{Deterministic scheme \cite{PIR} for $N=2$, $K=2$.}
    \label{eg2}
    \vspace*{-0.5cm}
\end{figure}

Another example with $N=2$ and $K=3$ is given in Fig.~\ref{eg2}. The size of a subpacket in this example is $N^K=8$, and the single subpackets of the three messages $W_1$, $W_2$ and $W_3$ are denoted by $(a_1,\dotsc,a_8)$, $(b_1,\dotsc,b_8)$ and $(c_1,\dotsc,c_8)$, respectively. The rate achieved in this example is $R=\frac{4}{7}$ for any message requirement, which equals the capacity in \eqref{cap} when $N=2$ and $K=3$.

The probabilistic approach is based on a set of universal queries that are used for all message requirements with equal probability, in the perspective of an individual database. There are two main probabilistic PIR schemes in the literature, proposed in \cite{leaky} and \cite{ChaoTian}, which are described next. For the example of $N=2$ and $K=2$, the set of possible queries sent to the two databases in the probabilistic approach from \cite{leaky} is given in Table~\ref{PPIR_N2K2}, where $W_1$ and $W_2$ represent the first and second messages, respectively. To retrieve any message, the user selects one query option from the query set, i.e., one row of Table~\ref{PPIR_N2K2}, with equal probability, and transmits the corresponding queries to the two databases. 

\begin{table}[h]
\begin{center}
\begin{tabular}{|c||c|c||c|c|}
\hline
& \multicolumn{2}{c||}{Retrieve $W_1$} & \multicolumn{2}{c|}{Retrieve $W_2$} \\
\hline
Prob. & DB $1$ & DB $2$ & DB $1$ & DB $2$\\ 
\hline
$\frac{1}{4}$ & $W_1$ & $-$ & $W_2$ & $-$\\
\hline
$\frac{1}{4}$ & $-$ & $W_1$ &  $-$ & $W_2$\\
\hline
$\frac{1}{4}$ & $W_2$ & $W_1\!+\!W_2$ & $W_1$ & $W_1\!+\!W_2$\\
\hline
$\frac{1}{4}$ & $W_1\!+\!W_2$ & $W_2$ & $W_1\!+\!W_2$ & $W_1$\\
\hline
\end{tabular}
\end{center}
\vspace{-0.5em}
\caption{Probabilistic scheme in \cite{leaky} for $N=2$, $K=2$.}
\label{PPIR_N2K2}
\vspace*{-0.2cm}
\end{table}

User privacy is guaranteed since each individual database will always receive one query from $4$ available query options $\{W_1,W_2,W_1\!+\!W_2,-\}$ with equal probability, irrespective of the user's message requirement (here, ``-'' corresponds to the ``empty'' query, i.e., ``no'' query). The rate achieved in this example for any message is,
\begin{align}\label{prob1rate}
    R=\left(\frac{\frac{1}{4}\times L+\frac{1}{4}\times L+\frac{1}{4}\times2L+\frac{1}{4}\times2L}{L}\right)^{-1} =\frac{2}{3},
\end{align}
where $L$ is the message length. The expression in \eqref{prob1rate} is obtained by considering the fact that the first two query options in Tables~\ref{PPIR_N2K2} require only $L$ symbols to be downloaded while the last two options require $2L$ symbols. The inverse of the expected download cost is calculated as the rate in \eqref{prob1rate}, which is the same as the capacity in \eqref{cap} for $N=2$ and $K=2$.

The probabilistic scheme proposed in \cite{ChaoTian} is based on the same principles as \cite{leaky}, except that it does not have the message symmetry present in \cite{leaky}. As an illustration of the scheme in \cite{ChaoTian}, consider the case of $N=2$ and $K=2$ again. As a reduced version of Table~\ref{PPIR_N2K2}, the explicit forms of the query options are provided in Table~\ref{PPIR2_N2K2}.

\begin{table}[h]
\begin{center}
\begin{tabular}{|c||c|c||c|c|}
\hline
& \multicolumn{2}{c||}{Retrieve $W_1$} & \multicolumn{2}{c|}{Retrieve $W_2$} \\
\hline
Prob. & DB $1$ & DB $2$ & DB $1$ & DB $2$\\ 
\hline
$\frac{1}{2}$ & $-$ & $W_1$ & $-$ & $W_2$\\
\hline
$\frac{1}{2}$ & $W_1\!+\!W_2$ & $W_2$ & $W_1\!+\!W_2$ & $W_1$\\
\hline
\end{tabular}
\end{center}
\vspace{-0.5em}
\caption{Probabilistic scheme in \cite{ChaoTian} for $N=2$, $K=2$.}
\label{PPIR2_N2K2}
\vspace*{-0.2cm}
\end{table}

To illustrate the general mechanism in \cite{ChaoTian}, consider an example with $N=3$ and $K=3$. Each subpacket in this scheme consists of $N-1=2$ symbols of each message. Let a single subpacket of the first, second and third message be denoted by $W_1=(a_1,a_2)$, $W_2=(b_1,b_2)$ and $W_3=(c_1,c_2)$, respectively. Assume that the user's required message is $W_2$. A dummy bit is appended to each subpacket in each message as, $W_1=(a_0,a_1,a_2)$, $W_2=(b_0,b_1,b_2)$ and $W_3=(c_0,c_1,c_2)$, where $a_0=b_0=c_0=0$. The first step  is to choose a random key of length $K-1=2$, from the set $\{0,\dotsc,N-1\}^{K-1}=\{0,1,2\}^2$. Assume that the chosen random key is $F=(0,2)$. Then, the query sent to database $n$ is of the form $Q_n=(\alpha,\beta,\gamma)$, where $\alpha=0$, $\gamma=2$, $\beta=(n-1-\sum_{i=1}^2 F(i))_N$, where $(\cdot)_N$ is the modulo $N$ operation. In other words, all elements except the second (required message index) in each length $K$ query vector are copied from the random key $F$ while the second element is chosen such that each query satisfies,
\begin{align}
    \left(\sum_{i=1}^3 Q_n(i)\right)_{\!\!N}=n-1,\quad n\in\{1,2,3\}.
\end{align}
The three query vectors sent to the three databases are,
\begin{align}
    Q_1=(0,1,2),\quad Q_2=(0,2,2), \quad Q_3=(0,0,2).
\end{align}
Once database $n$ receives the query $Q_n$, it calculates the answer as, $A_n=a_{Q_n(1)}+b_{Q_n(2)}+c_{Q_n(3)}$. The answers sent by the three databases are given by,
\begin{align}
    A_1=a_0\!+\!b_1\!+\!c_2, ~
    A_2=a_0\!+\!b_2\!+\!c_2, ~
    A_3=a_0\!+\!b_0\!+\!c_2,
\end{align}
from which the user can find the two bits of $W_2$ as,
\begin{align}
    b_1=A_1-A_3, \qquad 
    b_2=A_2-A_3,
\end{align}
since $b_0=0$. All possible random keys and the corresponding queries and answers of each database, when downloading $W_2$, are shown in Table~\ref{prob2}. 

\begin{table}[h]
\begin{center}
\scalebox{0.9}{
\begin{tabular}{ |p{0.3cm}||p{0.5cm}|p{1.5cm}||p{0.5cm}|p{1.5cm}||p{0.5cm}|p{1.5cm}|  }
 \hline
 $F$  &\multicolumn{2}{c||}{DB 1} & \multicolumn{2}{c||}{DB 2} & \multicolumn{2}{c|}{DB 3}\\
 \hline
 & \hspace{0.01em} $q_1$ & \hspace{1em} $A_1$ & \hspace{0.01em} $q_2$ & \hspace{1em} $A_2$ & \hspace{0.01em} $q_3$ & \hspace{1em} $A_3$\\
 \hline
 $00$ & $000$ & $a_0\!+\!b_0\!+\!c_0$ & $010$ & $a_0\!+\!b_1\!+\!c_0$ & $020$ & $a_0\!+\!b_2\!+\!c_0$\\
 \hline
 $10$ & $120$ & $a_1\!+\!b_2\!+\!c_0$ & $100$ & $a_1\!+\!b_0\!+\!c_0$ & $110$ & $a_1\!+\!b_1\!+\!c_0$\\
 \hline
 $20$ & $210$ & $a_2\!+\!b_1\!+\!c_0$ & $220$ & $a_2\!+\!b_2\!+\!c_0$ & $200$ & $a_2\!+\!b_0\!+\!c_0$\\
 \hline
 $01$ & $021$ & $a_0\!+\!b_2\!+\!c_1$ & $001$ & $a_0\!+\!b_0\!+\!c_1$ & $011$ & $a_0\!+\!b_1\!+\!c_1$\\
 \hline
 $11$ & $111$ & $a_1\!+\!b_1\!+\!c_1$ & $121$ & $a_1\!+\!b_2\!+\!c_1$ & $101$ & $a_1\!+\!b_0\!+\!c_1$\\
 \hline
 $21$ & $201$ & $a_2\!+\!b_0\!+\!c_1$ & $211$ & $a_2\!+\!b_1\!+\!c_1$ & $221$ & $a_2\!+\!b_2
 \!+\!c_1$\\
 \hline
 $02$ & $012$ & $a_0\!+\!b_1\!+\!c_2$ & $022$ & $a_0\!+\!b_2\!+\!c_2$ & $002$ & $a_0\!+\!b_0\!+\!c_2$\\
 \hline
 $12$ & $102$ & $a_1\!+\!b_0\!+\!c_2$ & $112$ & $a_1\!+\!b_1\!+\!c_2$ & $122$ & $a_1\!+\!b_2\!+\!c_2$\\
 \hline
 $22$ & $222$ & $a_2\!+\!b_2\!+\!c_2$ & $202$ & $a_2\!+\!b_0\!+\!c_2$ & $212$ & $a_2\!+\!b_1\!+\!c_2$\\
 \hline
\end{tabular}}
\end{center}
\vspace{-0.5em}
\caption{Probabilistic scheme in \cite{ChaoTian} for $N=3$, $K=3$ to download message $W_2$.}
\label{prob2}
\vspace*{-0.2cm}
\end{table}

Each random key is chosen with equal probability, and all except the first key requires the user to download one symbol each from all three databases while the first key requires the user to download a single symbol from only databases 2 and 3, since $a_0=b_0=c_0=0$ is globally known. Therefore, the rate achieved in this example is given by,
\begin{align}
    R=\left(\frac{\frac{1}{9}\times 2+\frac{1}{9}\times 3\times 8}{2}\right)^{-1}=\frac{9}{13},
\end{align}
which is exactly the capacity in \eqref{cap} when $N=3$, $K=3$. From the perspective of an individual database, each query from the possible sets of queries in Table~\ref{prob2} is received with equal probability irrespective of the user's message requirement, which guarantees user privacy. This can be seen by comparing each $q_i$ column in Tables~\ref{prob2} and~\ref{prob21} for downloading message $W_2$ and $W_1$, respectively. We left out the corresponding table for downloading message $W_3$ for space limitations. 

\begin{table}[h]
\begin{center}
\scalebox{0.9}{
\begin{tabular}{ |p{0.3cm}||p{0.5cm}|p{1.5cm}||p{0.5cm}|p{1.5cm}||p{0.5cm}|p{1.5cm}|  }
 \hline
 $F$  &\multicolumn{2}{c||}{DB 1} & \multicolumn{2}{c||}{DB 2} & \multicolumn{2}{c|}{DB 3}\\
 \hline
 & $q_1$ & \hspace{1em} $A_1$ & $q_2$ & \hspace{1em} $A_2$ &  $q_3$ & \hspace{1em} $A_3$\\
 \hline
 $00$ & $000$ & $a_0\!+\!b_0\!+\!c_0$ & $100$ & $a_1\!+\!b_0\!+\!c_0$ & $200$ & $a_2\!+\!b_0\!+\!c_0$\\
 \hline
 $10$ & $210$ & $a_2\!+\!b_1\!+\!c_0$ & $010$ & $a_0\!+\!b_1\!+\!c_0$ & $110$ & $a_1\!+\!b_1\!+\!c_0$\\
 \hline
 $20$ & $120$ & $a_1\!+\!b_2\!+\!c_0$ & $220$ & $a_2\!+\!b_2\!+\!c_0$ & $020$ & $a_0\!+\!b_2\!+\!c_0$\\
 \hline
 $01$ & $201$ & $a_2\!+\!b_0\!+\!c_1$ & $001$ & $a_0\!+\!b_0\!+\!c_1$ & $101$ & $a_1\!+\!b_0\!+\!c_1$\\
 \hline
 $11$ & $111$ & $a_1\!+\!b_1\!+\!c_1$ & $211$ & $a_2\!+\!b_1\!+\!c_1$ & $011$ & $a_0\!+\!b_1\!+\!c_1$\\
 \hline
 $21$ & $021$ & $a_0\!+\!b_2\!+\!c_1$ & $121$ & $a_1\!+\!b_2\!+\!c_1$ & $221$ & $a_2\!+\!b_2
 \!+\!c_1$\\
 \hline
 $02$ & $102$ & $a_1\!+\!b_0\!+\!c_2$ & $202$ & $a_2\!+\!b_0\!+\!c_2$ & $002$ & $a_0\!+\!b_0\!+\!c_2$\\
 \hline
 $12$ & $012$ & $a_0\!+\!b_1\!+\!c_2$ & $112$ & $a_1\!+\!b_1\!+\!c_2$ & $212$ & $a_2\!+\!b_1\!+\!c_2$\\
 \hline
 $22$ & $222$ & $a_2\!+\!b_2\!+\!c_2$ & $022$ & $a_0\!+\!b_2\!+\!c_2$ & $122$ & $a_1\!+\!b_2\!+\!c_2$\\
 \hline
\end{tabular}}
\end{center}
\caption{Probabilistic scheme in \cite{ChaoTian} for $N=3$, $K=3$ to download message $W_1$.}
\label{prob21}
\vspace*{-0.2cm}
\end{table}

Based on the PIR schemes discussed above, the suboptimality of the scheme in \cite{ori2} (shown in Fig.~\ref{eg1}) is caused by the inability of the users to explicitly download any non-required message symbols. Note that, the optimal scheme in \cite{PIR} (shown in Fig.~\ref{eg2}) lets the user download symbols from all messages explicitly, and use them as different pieces of side information, which allows the user to download multiple symbols of the required message privately. In contrast, the suboptimal scheme in \cite{ori2} only generates one piece of side information ($A_1$ in Fig.~\ref{eg1}), which limits the amount of required message bits that the user is allowed to download privately. However, the suboptimal scheme in \cite{ori2} preserves the privacy of messages that are not required by the user to a certain extent, by not allowing the user to download any non-required message bits explicitly, as opposed to the optimal schemes. We will see that this will come handy in symmetric privacy formulation next.

\section{SPIR Formulation and Schemes} \label{SPIR}

\begin{figure}[t]
    \centering
    \includegraphics[width=0.85\linewidth]{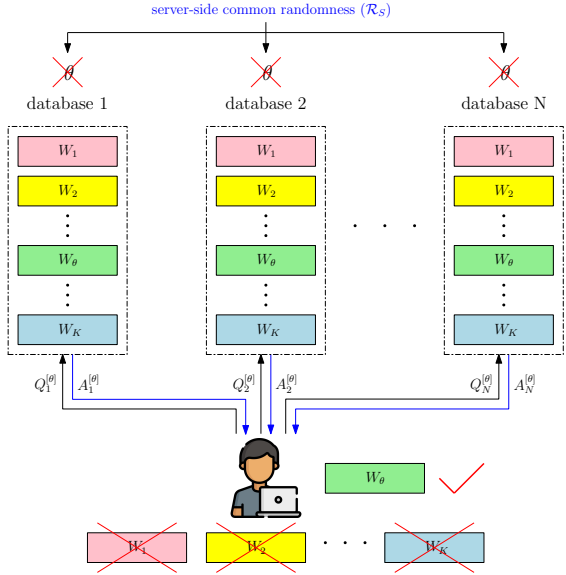}
    \caption{The system model of SPIR.}
    \label{SPIR_System_Model}
    \vspace*{-0.5cm}
\end{figure}

The concept of maintaining a two-way privacy requirement, where the databases are not allowed to learn the user's required message index while the user is not allowed to learn any information of the non-required messages is referred to as \emph{symmetric} PIR (SPIR) \cite{SPIR_ORI}, which is a non-trivial extension of PIR. In SPIR, the symmetry comes from the fact that the privacy of the user and the databases are desired to be guaranteed simultaneously; see Fig.~\ref{SPIR_System_Model}. The additional requirement that prohibits the user from learning anything beyond the required message is known as the database privacy constraint. It is proven that the user privacy constraint, database privacy constraint and  correctness constraint in SPIR jointly form a contradiction when no additional parameters are utilized \cite{SPIR_ORI,SPIR}. A well-known approach to perform SPIR is to introduce shared server-side common randomness at the databases in the server, that is unknown to the user.

\begin{figure}[t]
    \centering
    \includegraphics[width=0.8\linewidth]{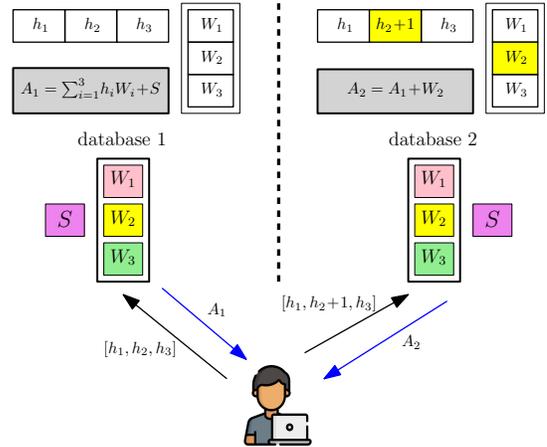}
    \caption{SPIR scheme in \cite{SPIR} for $N=2$ and $K=3$.}
    \label{SPIR_eg1}
    \vspace*{-0.5cm}
\end{figure}

The  SPIR scheme in \cite{SPIR} is constructed on the basis of the PIR schemes provided in \cite{original} and \cite{ori2} by appending extra server-side common randomness to the answers. Consider an SPIR example with $N = 2$ and $K = 3$, and assume that the user requires to download the second message $W_2$. The user sends thee randomly selected symbols from $\mathbb{F}_q$, denoted by $[h_1,h_2,h_3]$ to database 1, and $[h_1,h_2\!+\!1,h_3]$ to database 2. After receiving the query, each database calculates the dot product of the query and its own message set first, and adds a server-side common randomness symbol $S$ uniformly selected from $\mathbb{F}_q$ to the dot product. Then, each database transmits this information as answers back to the user. This process is shown in Fig.~\ref{SPIR_eg1}. User privacy and correctness constraints are satisfied the same way as in PIR (Section~\ref{pirschemes}). Database privacy is satisfied as the user is unable to learn any information about the other messages due to the existence of the unknown symbol $S$. The rate achieved in this example is $R=\frac{1}{2}$, which is the same as the rate of the example shown in Fig.~\ref{eg0}.

Following the generalized PIR scheme in \cite{ori2}, this example can be generalized to SPIR with arbitrary $K$ and $N$ by adding server-side common randomness to each answer. The corresponding rate is $R = 1 - \frac{1}{N}$ as in (\ref{spir}). Using the converse result in \cite{SPIR}, this rate is also the capacity of SPIR as long as sufficient server-side common randomness is available, i.e.,
\begin{align}
    C_{SPIR} = 1 - \frac{1}{N}.
\end{align}

Note that the capacity of SPIR does not depend on the number of messages $K$ and is strictly smaller than the capacity of PIR, i.e., $C_{SPIR} < C_{PIR}$, because of the additional database privacy constraint imposed. In addition, note that PIR capacity $C_{PIR}$ decreases with the number of messages $K$, and $C_{SPIR}=\lim_{K\rightarrow \infty} C_{PIR}$. Interestingly, it was shown in \cite{SPIR_atPIR} that the SPIR capacity can be increased to the PIR capacity if the user is able to pre-fetch a random subset of server-side common randomness (sufficient amount) from the server.

The SPIR scheme presented above is deterministic. Using the probabilistic PIR scheme in \cite{ChaoTian}, an alternative probabilistic SPIR approach is given in \cite{CommCost_ISIT2022}. For $N = 2$ and $K = 2$, the set of possible queries in this approach is given in Table~\ref{PSPIR_N2K2}, where $S$ is the server-side common randomness symbol only known by the databases. User privacy and correctness conditions are guaranteed the same way as in \cite{ChaoTian}, as described in Section~\ref{pirschemes}. Database privacy is satisfied due to the unknown symbol $S$. A general capacity-achieving SPIR scheme can be achieved by adding server-side common randomness to each query option in the generalized PIR scheme in \cite{ChaoTian}.

\begin{table}[h]
\begin{center}
\begin{tabular}{|c||c|c||c|c|}
\hline
& \multicolumn{2}{c||}{Retrieve $W_1$} & \multicolumn{2}{c|}{Retrieve $W_2$} \\
\hline
Prob. & DB $1$ & DB $2$ & DB $1$ & DB $2$\\ 
\hline
$\frac{1}{2}$ & $S$ & $W_1\!+\!S$ & $S$ & $W_2\!+\!S$\\
\hline
$\frac{1}{2}$ & $W_1\!+\!W_2\!+\!S$ & $W_2\!+\!S$ & $W_1\!+\!W_2\!+\!S$ & $W_1\!+\!S$\\
\hline
\end{tabular}
\end{center}
\vspace{-0.5em}
\caption{Probabilistic SPIR scheme \cite{CommCost_ISIT2022} for $N=2$, $K=2$.}
\label{PSPIR_N2K2}
\end{table}

\section{Systematic Extensions of PIR}

In this section, we briefly describe the problem formulations and capacity results of some variants of PIR.
\begin{enumerate}
    \item PIR with coded databases \cite{coded}: This problem considers $N$ non-colluding databases storing $K$ messages that are $(N,M)$ MDS coded. The capacity of coded PIR is,
    \begin{align}
        C_{coded}=\left(1+\frac{M}{N}+\cdots+\frac{M^{K-1}}{N^{K-1}}\right)^{-1},
    \end{align}
    which is a generalization of the capacity of classical PIR with replicated storage. Replication corresponds to the special case of $M=1$.
    \item PIR with colluding databases \cite{colluding}: This problem considers a system of $N$ databases where up to $T$ of them can collude. The capacity of colluded PIR is,
    \begin{align}
        C_{colluded}=\left(1+\frac{T}{N}+\cdots+\frac{T^{K-1}}{N^{K-1}}\right)^{-1},
    \end{align}
    which is equivalent to the capacity of classical PIR with $\frac{N}{T}$ non-colluding databases. 
    \item PIR with Byzantine and colluding databases \cite{byzantine}: This problem considers the presence of $B$ Byzantine databases out of the $N$ databases while any $T$ databases are allowed to collude. The capacity is given by, 
     \begin{align}
        C_{B}\!=\!\frac{N\!\!-\!\!2B}{N}\!\left(\!1\!+\!\frac{T}{N\!\!-\!\!2B}\!+\!\cdots\!+\!\frac{T^{K-1}}{(N\!\!-\!\!2B)^{K-1}}\!\right)^{-1} \!\!\!\!
    \end{align}
    which is equivalent to removing $2B$ databases out of the $N$ databases in a $T$-colluding setting. The scaling factor represents the fact that only $N-2B$ databases are of actual use even though all $N$ databases are accessed.
    \item Multi-message PIR (MM-PIR) \cite{MMPIR}: Here the user wants to download $P$ out of $K$ messages at a time, without revealing their identities to any of the $N$ databases. The capacity is given by,
    \begin{align}
        C_{MM-PIR}=\begin{cases}
        \frac{1}{1+\frac{K-P}{PN}}, & \text{if $P\geq\frac{K}{2}$}\\
        \frac{1-\frac{1}{N}}{1-\left(\frac{1}{N}\right)^{\frac{K}{P}}}, & \text{if $P\leq\frac{K}{2}$, $\frac{K}{P}\in\mathbb{N}$ }
        \end{cases}
    \end{align}
    The capacity when $P\leq\frac{K}{2}$ and $\frac{K}{P}\in\mathbb{N}$ is equal to the capacity of classical PIR with $\frac{K}{P}$ messages. This shows that using MM-PIR once is more efficient than using classical PIR $P$ times, to download $P$ messages. 
    \item Assymetric leaky PIR (AL-PIR) \cite{assymetric}: This variant of PIR studies the potential increase in the capacity, when a pre-determined amount of information is allowed to leak. AL-PIR considers both user and database privacy (SPIR), and proposes a scheme that performs under arbitrary information leakage budgets.
\end{enumerate}

\section{Conceptual Extensions of PIR}

\subsection{Private Set Intersection (PSI)}

\begin{figure}[t]
    \centering
    \includegraphics[width=0.8\linewidth]{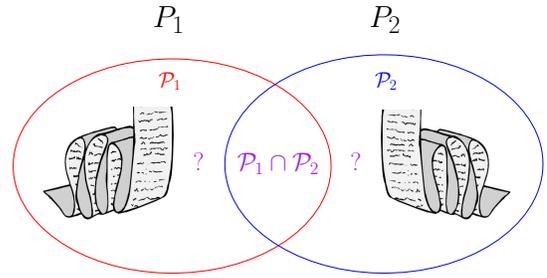}
    \caption{The system model for two-party PSI.}
    \label{PSI_System_Model}
    \vspace*{-0.4cm}
\end{figure}

Private set intersection (PSI) refers to the problem in which two parties $P_1$ and $P_2$ wish to determine the common elements jointly within their element sets without leaking any further information to each other about the remaining elements in their sets \cite{PSI_first}. The element sets associated with $P_1$ and $P_2$ are denoted by $\mathcal{P}_1$ and $\mathcal{P}_2$, which are selected from a global alphabet $\mathcal{A}$, based on an arbitrary statistical distribution. The basic two-party PSI system model is shown in Fig.~\ref{PSI_System_Model}.

The problem of PSI is motivated by practical security applications. For instance, consider a situation where an airline company has a list of its passengers while the national security agency (NSA) has a list of its suspected terrorists. NSA wants to check whether any of the terrorists is boarding a flight without revealing the entire list of terrorist suspects. Meanwhile, the airline company wishes to identify the terrorist suspects without revealing the entire list of its customers. Both parties are making an effort to determine the intersection of their respective lists in a private manner. Hence, three requirements are involved in the two-party PSI: First, at least one of the two parties should be able to decode the intersection correctly when the PSI process is complete. This requirement is called the PSI correctness constraint. Second and third, the privacy of the remaining elements in $\mathcal{P}_1$ and $\mathcal{P}_2$ must be guaranteed. These last two requirements are called the PSI $P_1$ and $P_2$ privacy constraints, respectively. 

As discussed in Section~\ref{SPIR}, there are three fundamental requirements in the SPIR problem, namely, correctness, user privacy, and database privacy. It was shown in \cite{PSI} that the three requirements of the PSI map one-to-one to the three requirements of the SPIR. Precisely, assuming without loss of generality that $P_1$ initiates the PSI process, $P_1$ privacy in PSI corresponds to user privacy in SPIR; $P_2$ privacy in PSI corresponds to database privacy in SPIR; and correctness in determining the set intersection in PSI corresponds to correctness in decoding in SPIR. Here, note that $P_1$ wishes to learn whether its own elements are also in $P_2$ without revealing the identity of its own elements (user privacy) and without learning anything further than the existence of those particular elements in $P_2$ (database privacy). Further, note that, since $P_1$ has multiple elements, it will want to check the existence of these multiple elements all at once, considering the fact that multi-message SPIR (MM-SPIR) may be more efficient than multiple application of single-message SPIR, as it happened in the case of multi-message PIR. Therefore, PSI is exactly equivalent to MM-SPIR. As a result, all known converse results as well as the existing achievable schemes for MM-SPIR translate directly into PSI. 

Finally, the PSI problem (and the equivalent MM-SPIR) can be extended to information theoretic secure multi-party PSI (MP-PSI), where multiple (more than two) parties wish to jointly determine the intersection of their respective element sets while protecting the privacy of their remaining elements \cite{MP-PSI}. This extension is achieved in \cite{MP-PSI} via the introduction of an intricate common randomness distribution scheme among the multiple parties before the MP-PSI process starts.

\subsection{Private Set Union (PSU)}

Following the discussion in the last subsection, as a dual problem of PSI, private set union (PSU) refers to the problem in which two parties aim to compute the union of their element sets jointly without revealing anything beyond the union to each other \cite{PSU}. Similar to the two-party PSI problem, there are three underlying requirements in the two-party PSU problem formulation: At least one party should be able to obtain the union without any error; this is called the PSU correctness constraint. The privacy of the remaining elements in $P_1$ needs to be kept against $P_2$; this is the PSU $P_1$ privacy constraint. The privacy of the remaining elements in $P_2$ needs to be kept against $P_1$; this is the PSU $P_2$ privacy constraint. The duality between PSU and PSI can be seen through the De Morgan's law: $\overline{A \cup B} = \overline{A} \cap \overline{B}$. Thus,  we have $A \cup B = \overline{\overline{A} \cap \overline{B}}$, which implies that the set union can be found using a combination of set intersection and set complement. Therefore, by mapping SPIR correctness constraint, user privacy constraint and database privacy constraint to PSU correctness constraint, $P_1$ privacy constraint and $P_2$ privacy constraint respectively, the equivalence of PSU and SPIR is established in \cite{FSL-PSU}, which says that PSU is equivalent to MM-SPIR. Moreover, by using the scheme in \cite{MP-PSI} for reference, \cite{PSU} shows that basic two-party PSU can be generalized to multi-party PSU (MP-PSU) in an information-theoretic secure sense.

\subsection{PSU-based Federated Submodel Learning (FSL)}\label{psu_fsl}

Federated learning (FL)\cite{FL2} is a framework where a central machine learning model is collectively trained by a large number of clients, using their local data. In FL, each client downloads the entire model, updates it, and uploads the updates back to the central server. This process is inefficient in terms of the communication and computation overhead, since each client downloads and uploads the entire model even if the data available at a given client is not sufficient to train the entire model. One solution to this problem is federated submodel learning (FSL) \cite{billion,secureFSL,paper1}, where the central learning model is divided into multiple submodels based on different types of training data, as this allows each client to only download and update the submodel(s) relevant to the client's local data. However, the submodel indices requested by a client and the values of updates uploaded leak information about the client's private data. Hence, to ensure the privacy of the client's local data in FSL, the following two questions need to be answered. How can the clients download the desired submodels without revealing their indices to the server that stores the submodels? How can the clients update the desired submodels without revealing their indices or the values of the updates to the server? The first question refers to a \emph{private read} problem which is the same as PIR, and the second question refers to a \emph{private write} problem which is an important conceptual extension of PIR.

In classical FL, secure aggregation is used to guarantee privacy of the clients' data, in which the server can only learn the aggregate model from the clients and nothing beyond that \cite{PracticalSecureAgg}. As discussed in the last subsection, in MP-PSU, if one of the parties is selected as a leader party to derive the union, the MP-PSU privacy constraints require that this leader party can obtain only the union from the remaining parties and nothing beyond that. Therefore, \cite{FSL-PSU} proposes a new private FSL approach which is based on PSU and SPIR. The core idea behind this scheme is to preserve client-side privacy by not allowing the server to learn any information beyond the subset of submodels (or the aggregation of updates of these submodels) collectively updated by the clients, i.e., the server learns nothing about any individual client's contribution to the ultimate result. Thus, no extra information about the clients' local training data is leaked to the server. 

The sketch of this scheme is as follows. Within the initialization stage, two independent databases in the server store the replicated storage of all submodels and a sufficient amount of server-side common randomness.\footnote{The case of two databases is considered here, and the achievable scheme works for any number of databases with minor modifications.} Initially, the two databases distribute/duplicate client-side common randomness to each selected client using random SPIR \cite{RSPIR} and classical one-time pads. Next, the clients utilize the established client-side common randomness to have the server reliably decode the union of indices of submodels to be updated collectively by the clients in a private manner. This phase is referred to as the FSL-PSU-read phase. Then, the two databases broadcast the current versions of the submodels in the set union to the clients. The clients update the desired submodels using their local training data. Finally, the clients use a variation of FSL-PSU to write the updates back to the databases privately. This phase basically performs secure aggregation, and is referred to as the PSU-write phase. 

To improve the communication efficiency, instead of enforcing each client to send the same answer to both databases, all the selected clients are divided into two groups such that most of them merely communicate with the database they belong to. A very small subset of clients is randomly selected as intermediators to route the information received by the two databases from their associated clients, to compensate for the absence of communication between the two databases. By utilizing the server-side and client-side common randomness carefully, the private FSL scheme in \cite{FSL-PSU} is shown to be robust against client drop-outs, client late-arrivals and database drop-outs. Client drop-outs occur when some clients leave during the training process, and client late-arrivals occur when some clients come up with late answers that may leak additional information about these late clients to the databases because of the wrong judgement made by the server that these clients have already dropped-out. Database drop-outs occur when databases do not function smoothly at a given time instance.

\begin{figure*}
    \centering
    \begin{subfigure}[b]{0.4\textwidth}
        \centering
        \includegraphics[width=\textwidth]{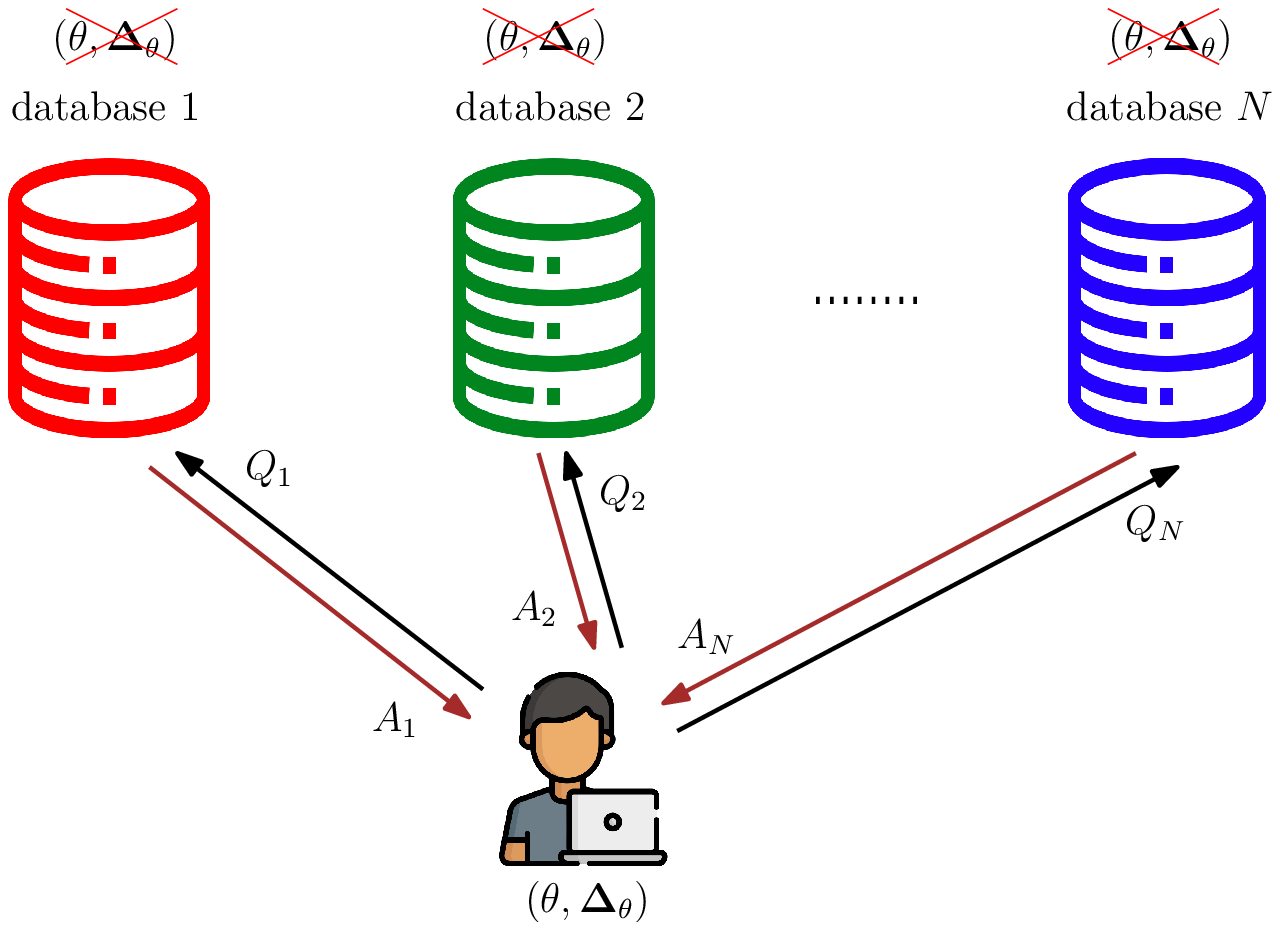}
        \caption{Reading phase.}
        \label{read}
    \end{subfigure}
    \hfill
    \begin{subfigure}[b]{0.4\textwidth}
        \centering
        \includegraphics[width=\textwidth]{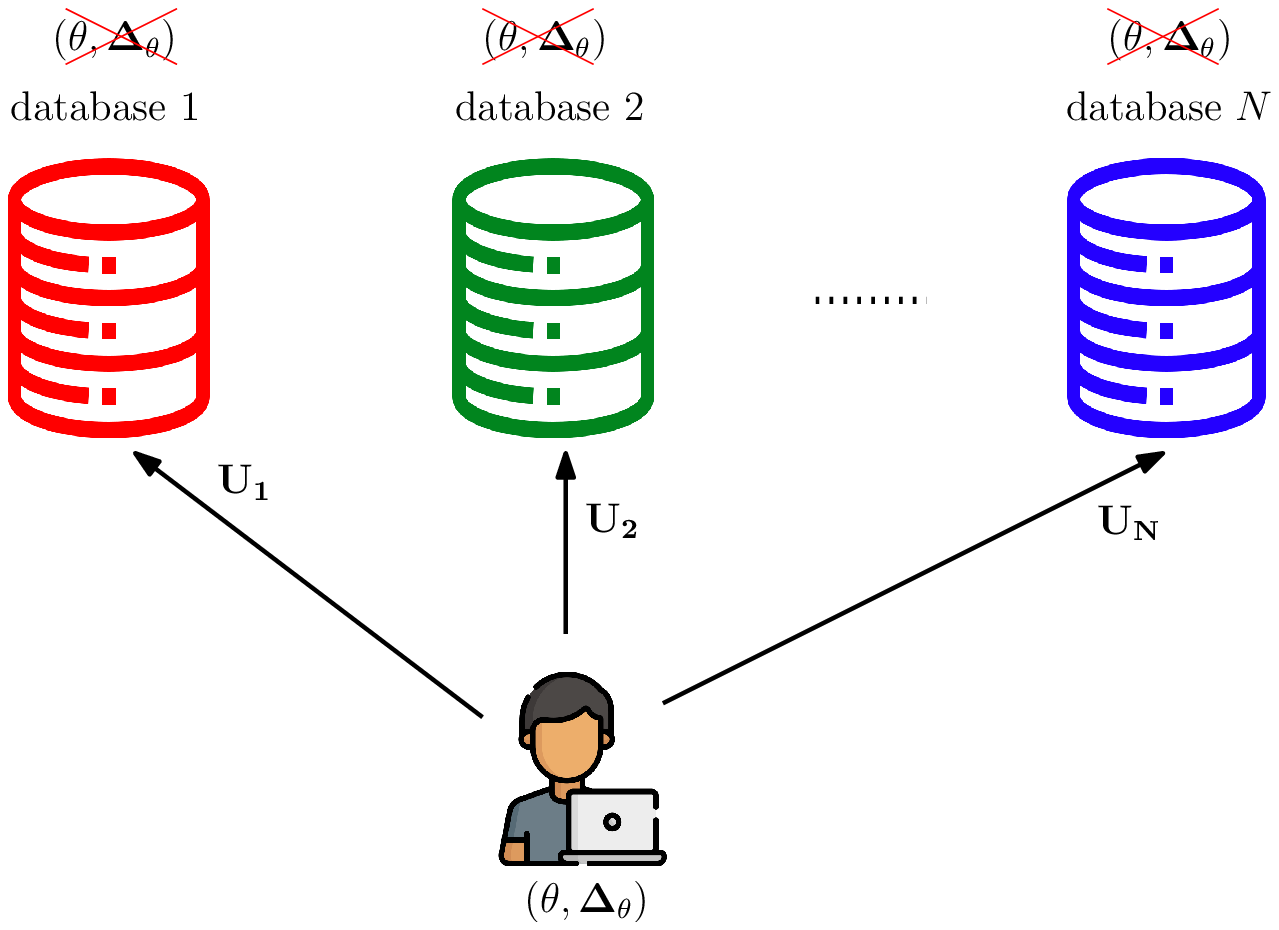}
        \caption{Writing phase.}
        \label{write}
    \end{subfigure}
    \caption{System model of private FSL with PRUW.}
    \label{pruwmodel}
    \vspace*{-0.4cm}
\end{figure*}

\subsection{Private Read-Update-Write (PRUW)}

Private read-update-write (PRUW)\cite{pruw_journal,dropout,ourICC,rw_jafar} is the concept where a user downloads, updates and uploads the updates back to a chosen section of a data storage system, without revealing the content downloaded, uploaded or their positions, e.g., updated section index. PRUW is essentially the combination of \emph{private reading} and \emph{private writing} described previously. PRUW has two main applications in distributed learning, namely, private FSL and private FL with sparsification.

Private FSL, as discussed before, requires a client to download a required submodel, update it and upload the updates back to the submodel without revealing the submodel index or the values of updates. The method proposed in \cite{FSL-PSU} achieves it by preventing the databases from learning any information beyond the union of submodels downloaded/updated by all clients. PRUW achieves stronger privacy guarantees in FSL by leaking zero information about any of the submodel indices or values of updates, updated by any individual or collection of clients. In particular, at any given time instance of the FSL process, the server has zero information on the submodels that have been updated so far, or the values of the updates.

The other application of PRUW is private FL with sparsification. Gradient sparsification \cite{sparse1,GGS,adaptive,conv,overtheair,rtopk,timecorr} is a mechanism used in most learning tasks to reduce the communication overhead. In gradient sparsification, each client only uploads a selected set of updates (e.g., most significant, randomly chosen, etc.), along with their indices to the server to reduce the upload cost. In cases where the updates are selected based on their significance, the indices of the sparse updates leak information about the types of data that the client has. Moreover, it has been shown in \cite{MembershipInterference, featureLeakage, InvertingGradients, DeepLeakage, SecretSharer,BeyondClassRepresentatives, comprehensive} that the values of the updates can be used to infer information about the client's local data. This problem is solved by PRUW, which facilitates private communication of the sparse updates and parameters, by hiding their values and indices from the servers.

Next we describe how information-theoretic privacy is achieved by PRUW in FSL and FL with top $r$ sparsification, followed by some other variants of PRUW.

\subsubsection{PRUW in FSL} \label{pruw_fsl}

As explained in Section~\ref{psu_fsl}, private FSL refers to the problem where a client reads (downloads), updates and writes (uploads) the updates back to a desired submodel out of multiple submodels in the FSL model, without revealing the submodel index or the values of updates. The two phases in the FSL process, where the client reads the parameters of the required submodel and writes the updates back to the submodel are known as the reading phase and the writing phase, respectively. Formally, the privacy constraint on the submodel index requires the mutual information between the desired submodel index $\theta$ and all the information sent by the client in both reading and writing phases at all time instances to be equal to zero, in the perspective of any individual database. Similarly, the privacy constraint on the values of updates requires the mutual information between the values of updates and all the information sent by the client at all time instances to be equal to zero. Moreover, there exists a security constraint, which prevents each individual database from learning any information about the submodels from its stored content. This is required for the privacy of the values of updates, since the changes in submodels at distinct time instances directly reveal information about the values of updates. The correctness constraint in the reading phase is the same as that of PIR, while the writing phase at time $t$ requires the updating submodel at time $t-1$ to be added with the corresponding updates, while all other submodels remain the same. The reading and writing costs are defined as the total number of bits downloaded and uploaded, normalized by the size of a submodel. The total cost is the sum of the reading and writing costs.

To solve the basic private FSL problem, \cite{rw_jafar,dropout,ourICC,pruw_journal} use the system model shown in Fig.~\ref{pruwmodel}, consisting of $N$ non-colluding databases, each storing $M$ independent submodels consisting of symbols from a large enough finite field $\mathbb{F}_q$. Similar to PIR, each client sends queries to databases to download the required submodel in the reading phase, for which the databases send answers, as shown in Fig.~\ref{read}. In the writing phase, the client sends updates to all databases, which are added to the relevant submodel parameters, without revealing the submodel index $\theta$ or the values of updates $\Delta_{\theta}$, see Fig.~\ref{write}. All schemes presented in \cite{rw_jafar,dropout,ourICC,pruw_journal} are based on the concept of cross subspace alignment \cite{CSA}. These schemes are designed along the lines of private distributed coded computing, where the submodels, queries and updates are secured by adding random noise from the finite field in a specific way, which essentially makes them random noise variables from the perspective of an individual database, based on Shannon's one-time pad theorem \cite{OTP}. Shanon's one-time pad theorem basically states that if $X$ is a random variable with any arbitrary distribution, and $Y$ is an independent and uniformly distributed random variable, both in the same finite field $\mathbb{F}_q$, the random variable  $Z=X+Y$ is uniformly distributed, and is independent of $X$. 

\begin{figure}
    \centering
    \includegraphics[width=0.6\columnwidth]{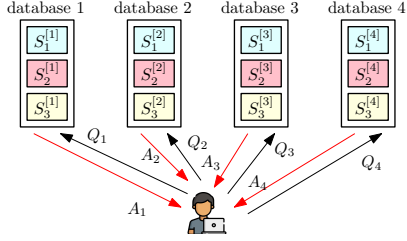}
    \caption{PRUW in FSL: Reading phase.}
    \label{mechanism}
    \vspace*{-0.4cm}
\end{figure}

The basic private FSL scheme proposed in \cite{pruw_journal,ourICC} adds random noise to all types of information that need to be kept private (submodel index, updates, submodels in storage), to guarantee information-theoretic privacy. This noise addition is done in terms of a carefully designed polynomial, and what is stored/sent to each database is a unique evaluation of this noise added polynomial. Each individual polynomial evaluation sent to a given database is simply random noise, while the collection of all $N$ evaluations reveals the hidden value. In other words, the storage, queries and updates corresponding to each database are shares of secrets\cite{ss}. All computations in both reading and writing phases are carried out on these noisy polynomial evaluations at each database, which are designed in such a way that all the noise components resulting from the computations are aligned along a specific subspace in an $N$ dimensional space, while the data components are aligned along specific directions that are linearly independent of the noise subspace. This makes it possible for a client to download the required submodel by only accessing the specific direction allocated to it in the $N$ dimensional space in the reading phase, and to write the real updates back to the specific direction, while also writing additional noise elements into the noise subspace, to guarantee privacy in the writing phase. 

As an example, consider an FSL system where four non-colluding databases store three submodels, denoted by $W_1$, $W_2$ and $W_3$. The three submodels are \emph{secrets} that need to be hidden by the databases that store them. Therefore, they must be stored as shares of secrets in the four databases. Let the secret share corresponding to $W_k$, stored in database $n$ be denoted by $S_k^{[n]}$, as shown in Fig.~\ref{mechanism}. The explicit form of $S_k^{[n]}$ is given by,
\begin{align}\label{poly}
    S_k^{[n]}&=w_k+(f-\alpha_n)(Z_k+\alpha_n Y_k)
\end{align}
where $w_k$ denotes a single bit of $W_k$. $f$ and $\alpha_n$ are globally known distinct constants and $Z_k$, $Y_k$ are random noise bits from $\mathbb{F}_q$. Note that what is stored in database $n$ is the evaluation of \eqref{poly} at the corresponding $\alpha_n$,\footnote{Each database has a distinct $\alpha_n$ associated with it.} which is basically random noise due to the added noise component, from Shannon's one-time pad theorem. Assume that the client wants to download $W_2$. In the reading phase, the client sends queries to the four databases which are also shares of the secret that contains the required submodel index. The secret, i.e., submodel index 2, is represented by the vector $[0 \ 1 \ 0]^T$, since there are only three submodels, and the query sent to database $n$ is given by,
\begin{align}\label{qu}
    Q_n=     
    \frac{1}{f-\alpha_n}\begin{bmatrix}0\\1\\0 \end{bmatrix}+\begin{bmatrix}\Bar{Z}_1\\\Bar{Z}_2\\\Bar{Z}_3 \end{bmatrix},
\end{align}
where $\bar{Z}_i$ are random noise bits. The evaluation of \eqref{qu} at the respective $\alpha_n$ is sent to each database. These are also random noise vectors due to the added noise, which guarantees the privacy of the required submodel index. Once the databases receive the queries from the client, they calculate the answers as the dot products of the storage and the received queries. The explicit form of the answer calculated by database $n$ is,
\begin{align}\label{ans}
    A_n=\frac{1}{f-\alpha_n}w_{2}+V_0+\alpha_n V_1+\alpha_n^2 V_2,
\end{align}
where $V_0,V_1,V_2$ are obtained by combining all constant terms and coefficients of $\alpha_n^i$ terms for $i=1,2$, respectively, of the dot product. The four answers obtained by the four databases, i.e., evaluation of \eqref{ans} at four distinct $\alpha_n$s are used to retrieve $w_{2}$, along with $V_0$, $V_1$ and $V_2$. 

\begin{figure}
    \centering
    \includegraphics[width=0.5\textwidth]{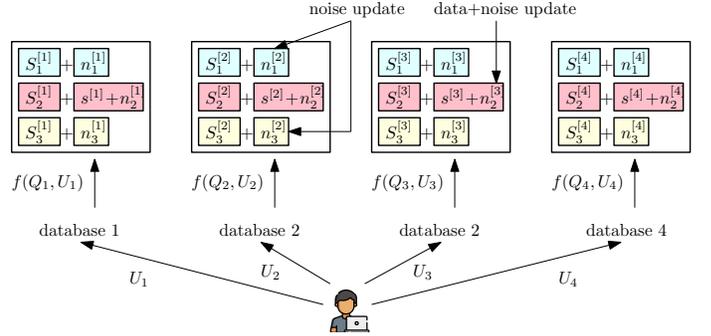}
    \caption{PRUW in FSL: Writing phase.}
    \label{wr}
    \vspace*{-0.4cm}
\end{figure}

In the writing phase, the client sends the update of $w_{2}$ denoted by $\Delta_{2}$ to the four databases as the evaluations of,
\begin{align}\label{upd}
    U_n&=\Delta_{2}+(f-\alpha_n)\dot{Z},
\end{align}
where $\dot{Z}$ is a random noise bit. The evaluations of \eqref{upd} sent to the databases are simply random noise bits due to the added random noise, which guarantees the privacy of the update $\Delta_{2}$. At each database, the received noisy update is placed at the relevant position, i.e., at the second row in Fig.~\ref{mechanism}, with the aid of the query received in the reading phase as,
\begin{align}\label{incr}
    \Bar{U}_n=(f-\alpha_n)\times U_n\times Q_n,
\end{align}
which is the incremental update that is added to the existing storage in each database. In this process, the noise component of the shares of secrets corresponding to $W_1$ and $W_3$, stored in each database get updated while both the data and noise components get updated in the secret shares corresponding to $W_2$ in all databases. This is shown in Fig.~\ref{wr}. 

The efficiency of PRUW in FSL increases with the number of databases by allowing multiple updates to be combined to a single bit in the writing phase, which significantly reduces the writing cost. With large enough number of databases, PRUW in FSL can be performed by only downloading and uploading twice as many bits as the size of a submodel.

\subsubsection{PRUW in FL With Top $r$ Sparsification} 

\begin{figure}
    \centering
    \includegraphics[width=0.85\columnwidth]{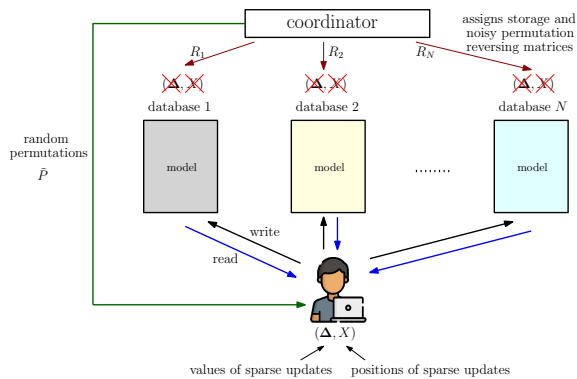}
    \caption{PRUW in FL with sparsification.}
    \label{sparsefl}
    \vspace*{-0.5cm}
\end{figure}

In FL with top $r$ sparsification, clients only download and upload the most significant $r'$ and $r$ fractions of parameters and updates, respectively, to reduce the communication cost. Clients typically send the sparse updates along with their indices to databases when privacy is not a concern. However, as discussed before, revealing the values and the indices of the sparse updates compromise the client's privacy. Motivated by these privacy concerns, the problem of PRUW in FL with top $r$ sparsification is introduced in \cite{pruw_journal,model,rps}. The privacy constraint on the downloaded (uploaded) parameter (update) indices requires the mutual information between the real indices and all the information received by an individual database to be equal to zero. Privacy on the values of updates and security of the model parameters are the same as in PRUW in FSL. The correctness constraint in the reading phase requires the clients at time $t$ to download the most commonly updated $r'$ fraction of parameters at time $t-1$, without revealing their indices, and the correctness constraint in the writing phase requires the sparse updates to be added to the respective real indices of the model while the rest of the parameters remain the same. The system model is shown in Fig.~\ref{sparsefl}, for the communication between a single client and $N$ non-colluding databases. 

PRUW facilitates the sparse read-write process without revealing the values or the indices of the sparse updates \cite{model,rps}. The values of the sparse updates are protected by adding random noise, as explained in Section~\ref{pruw_fsl}. The privacy of the indices of sparse updates is achieved by considering a random permutation of all parameters of the model which is only known by the clients. The clients send the indices of sparse updates using this permuted order. In order to rearrange the updates in the correct order at the databases for correctness, a noise added permutation reversing matrix is stored at each database, corresponding to the chosen random permutation, from which no information on the underlying permutation can be learned by the databases. The noise added random permutation matrix for database $n$ is denoted by $R_n$ in Fig.~\ref{sparsefl}.

The random permutation selection and the placement of the noisy permutation reversing matrix at each database is done at the initialization stage, before the learning process begins, with the help of a third party called the ``coordinator" as shown in Fig.~\ref{sparsefl}, which is also used in general PRUW to initialize the storage at each database prior to the learning process. The noisy permutation reversing matrix, along with the permuted sparse update indices are used to privately rearrange the sparse updates in the correct order, so that the updates are correctly added to the respective parameters.

\begin{figure}
    \centering
    \includegraphics[width=0.73\columnwidth]{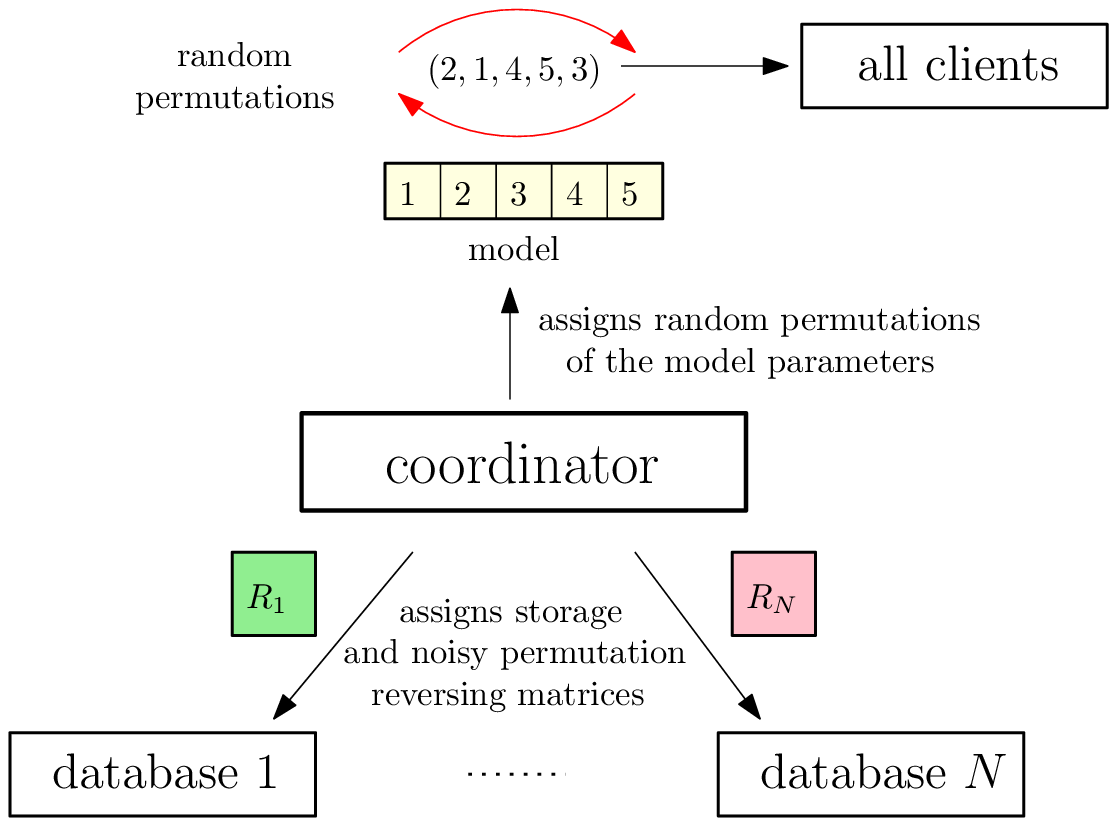}
    \caption{The private permutation mechanism.}
    \label{sparseg}
    \vspace*{-0.4cm}
\end{figure}

Consider a conceptual example where an FL model consists of five parameters as shown in Fig.~\ref{sparseg}. At the initialization stage, the coordinator sends a random permutation of the five parameters to all the clients involved in the learning process. Assume that this randomly chosen permutation is given by $\Tilde{P}=(2,1,4,5,3)$. At the same time, the database stores a noise added permutation-reversing matrix at each database. These matrices are also shares of a secret from which the databases are unable to learn anything about the underlying permutation. The noise added permutation-reversing matrix stored at database $n$ is of the form,\footnote{The exact noisy permutation-reversing matrix is a scaled version of \eqref{perm_rev}.}
\begin{align}\label{perm_rev}
    R_n=\begin{bmatrix}
        0 & 1 & 0 & 0 & 0\\
        1 & 0 & 0 & 0 & 0\\
        0 & 0 & 0 & 0 & 1\\
        0 & 0 & 1 & 0 & 0\\
        0 & 0 & 0 & 1 & 0\\
    \end{bmatrix}+\alpha_n X,
\end{align}
where $X$ is a random noise matrix of size $5\times5$. Note that the first part of \eqref{perm_rev} is essentially the permutation reversing matrix corresponding to $\Tilde{P}$, which is hidden from the databases by the added noise component, which ensures the privacy of the random permutation. 

To illustrate how the updates are uploaded privately without revealing their values or updates, consider an example where a client wants to update parameters $2$ and $3$. Then, the client sends the noise added updates of the form \eqref{upd} along with their permuted indices $1$ and $5$, since the real indices $2$ and $3$ are positioned at the first and fifth positions in $\Tilde{P}$. Note that the values of the updates are protected by the noise added in \eqref{upd}. The privacy of the sparse indices $2$ and $3$ is guaranteed since the permutation $\Tilde{P}$ is randomly chosen. Once each database receives the two updates $U_2$ and $U_3$ (corresponding to the real indices $2$ and $3$) along with the permuted indices $1$ and $5$, it calculates the privately rearranged updates by multiplying the noise added permutation-reversing matrix of the form \eqref{perm_rev} by $[U_2 \ 0 \ 0 \ 0 \ U_3]^T$. Note that this multiplication results in ``$[0\ U_2\ U_3\ 0\ 0]^T+$noise'', where the two updates corresponding to real indices $2$ and $3$ are correctly rearranged at the respective positions, while leaking no information about the real indices. Privately rearranged updates are  added to the existing storage. 

In the reading phase, the databases choose the most popular $r'$ fraction of (permuted) indices received by the clients at the previous time instance, and send them to the clients. Note that the databases are still unaware of the corresponding real indices since the selection is done based on the permuted indices. Each client obtains the real indices corresponding to the received permuted indices using the known random permutation $\Tilde{P}$. Now, for the client to access the real parameters corresponding to the permuted indices chosen by the databases, each database calculates the dot product between its storage, i.e., the five noise-added parameters in the Fig.~\ref{sparseg}, and the column of the noise added permutation-reversing matrix corresponding to each chosen permuted index, and sends it to the client. The client obtains the parameters from the answers received by all databases. Note that private FSL with top $r$ sparsification \cite{pruw_journal,sparse1} can be performed by combining the above concepts with the scheme explained in Section~\ref{pruw_fsl}. The resulting reading and writing costs of both private FL and FSL with top $r$ sparsification when $N$ is large, is approximately $2r'$ and $2r$, respectively, which is significantly small since $r',r\approx 10^{-2}$ in practice.

\begin{figure}[t]
    \centering
    \begin{subfigure}[b]{0.5\textwidth}
        \centering
        \includegraphics[width=\textwidth]{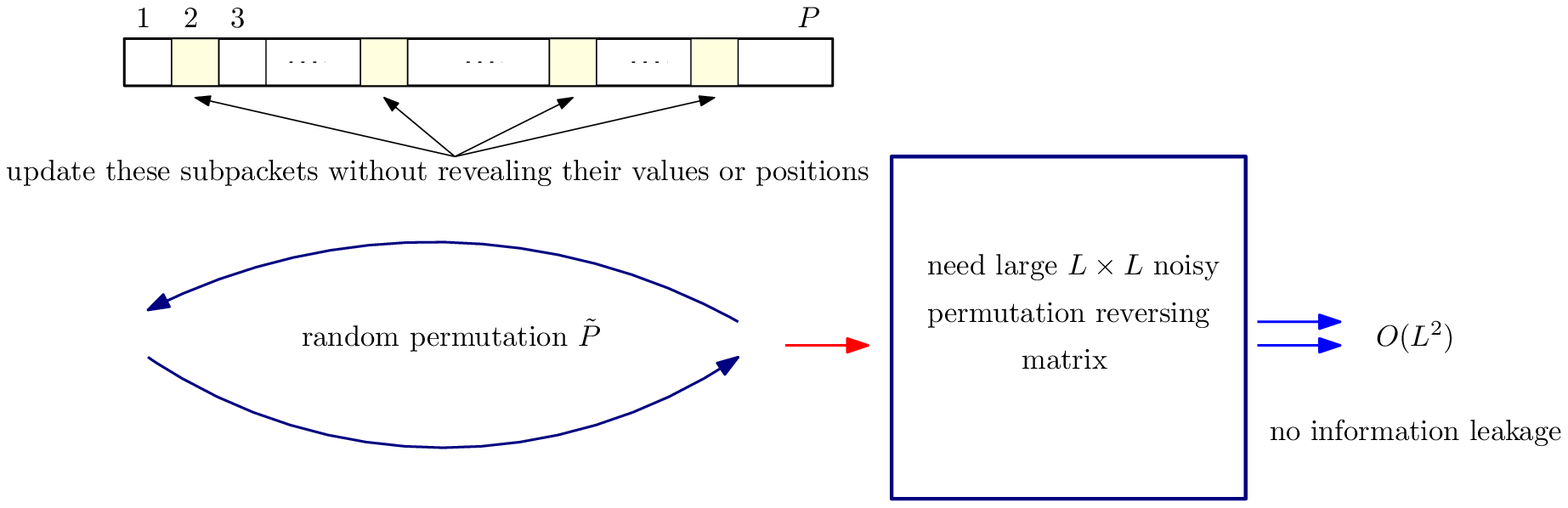}
        \caption{Without segmentation.}
        \label{prob}
    \end{subfigure}
    \vfill \vspace*{0.5cm}
    \begin{subfigure}[b]{0.5\textwidth}
        \centering
        \includegraphics[width=\textwidth]{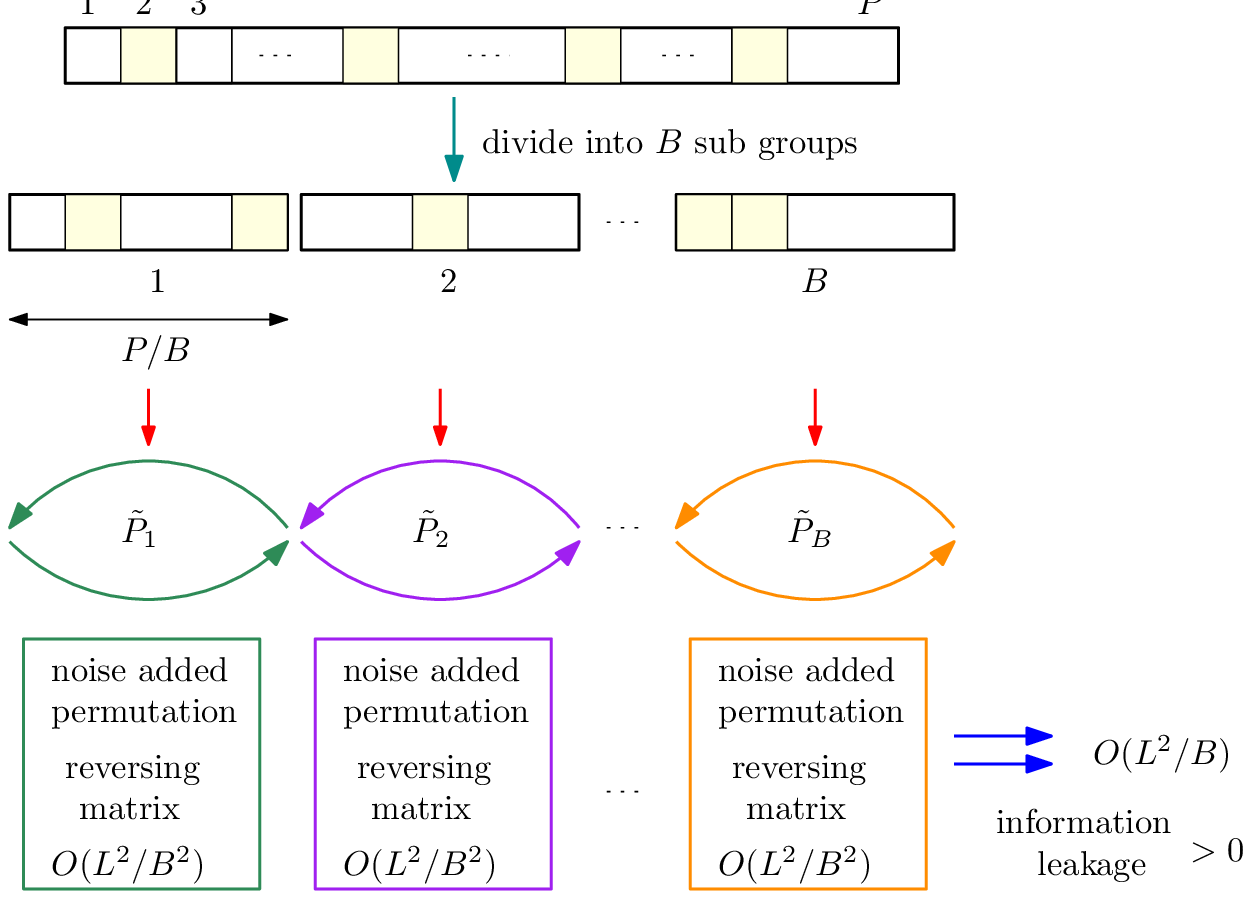}
        \caption{With segmentation.}
        \label{sol}
    \end{subfigure}
    \caption{Segmentation in permutation techniques.}
    \label{idea}
    \vspace*{-0.4cm}
\end{figure}

A main drawback of this process is the large storage cost caused by the significantly large noise-added permutation reversing matrices, which increases with the number of parameters in the FL model. The size of the noise added permutation-reversing matrices can be reduced at the expense of a certain amount of information leakage. This is achieved by dividing the FL model into multiple segments and carrying out permutations in each segment separately \cite{rps,model,weak_private_journal}. This idea is illustrated in Fig.~\ref{idea}.

When performing permutations in each segment separately as shown in Fig.~\ref{sol}, the client sends the permuted indices of the sparse updates of each segment separately to the databases. This leaks information about the indices of the sparse updates, since the databases learn how the $r$ fraction of updates are distributed among the segments, i.e., the number of sparse updates in each segment. The information leakage on the indices of the sparse updates is defined by the mutual information between the real indices of the sparse updates and all the information sent by the client to any individual database. Note that no information is leaked on the values of updates since the same method of adding random noise in \eqref{upd} is used to hide the values of updates. With $B$ segments, the information leakage is characterized by the joint entropy of the $B$ random variables representing the numbers of sparse updates in the $B$ segments. Therefore, the information leakage and the storage cost of $O(\frac{L^2}{B})$, Fig.~\ref{idea}, are inversely proportional.

\begin{figure}
    \centering
    \includegraphics[width=0.45\textwidth]{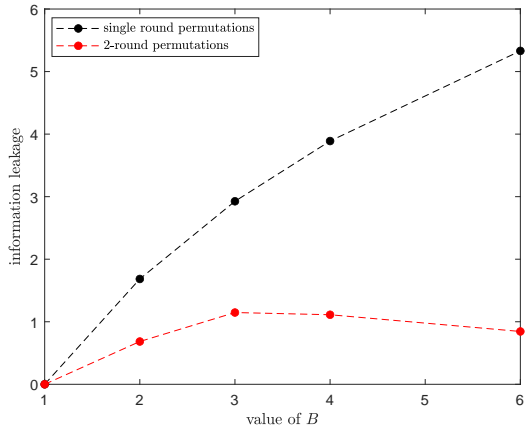}
    \caption{Information leakage with varying number of segments.}
    \label{grph}
    \vspace*{-0.4cm}
\end{figure}

The information leakage can be reduced further, by carrying out another stage of permutations. In addition to the $B$ within segment permutations shown in Fig.~\ref{idea}, another round of permutations can be performed among the segments. While this method also leaks the number of sparse updates in each segment, it does not reveal the mapping between the number of updates and the segment index. Therefore, the information leakage is reduced to the joint entropy of the set of $B$ random variables representing the distinct combinations of the numbers of sparse updates in the $B$ segments, irrespective of the order. Fig.~\ref{grph} shows the amounts of information leaked in the two cases (single-stage and two-stage permutations) for different number of segments in an example where the FL model consists of $12$ subpackets. The read-write costs of both cases are not significantly affected by segmentation, and remain the same at $\approx 2r'$ and $\approx 2r$.

\subsubsection{Rate-Distortion Trade Off in PRUW}

Although the communication cost of a private distributed learning process is significantly reduced by top $r$ sparsification, the storage cost is high due to the noisy permutation reversing matrices, when using PRUW to guarantee privacy. By allowing a pre-determined amount of distortion in the learning process, PRUW can be used at a lower storage cost as well as a lower communication cost. This mechanism does not require any permutations, and reduces the communication cost by not downloading (uploading) a randomly selected set of parameters (updates) in the reading (writing) phase. This method can also be thought of as random sparsification. Note that the ignored parameters and updates in the reading and writing phases result in a certain amount of distortion. In other words, this method performs PRUW in distributed learning while achieving a lower communication cost, at the expense of a given amount of distortion in the reading and writing phases. The rate-distortion trade off in PRUW is quantified by analyzing the minimum communication cost at different amounts of allowed distortion. While the system model, privacy, security and correctness constraints are the same as before, this problem setting defines distortion budgets for the reading and writing phases. The goal is to achieve the minimum communication cost in the private read-write process while guaranteeing information-theoretic privacy of client's local data and maintaining the distortion under the allowed budget. The distortion in reading (writing) phase is defined as the total number of parameters (updates) incorrectly downloaded (uploaded), normalized by the size of the model/submodel. It is shown in \cite{pruw_journal,rd} that the rate distortion trade off in PRUW is linear, i.e., the reading and writing costs are given by,
\begin{align}
    C_R=(1-D_R)C_1,\quad C_W=(1-D_W)C_2,
\end{align}
where $C_R$, $C_W$ represent the reading and writing costs, $D_R$, $D_W$ represent the distortion budgets in reading and writing phases, and $C_1$, $C_2$ are the reading and writing costs achieved by the basic PRUW without distortion. 

Note that PRUW in FL with top $r$ sparsification also introduces some amount of distortion. However, top $r$ sparsification selects the most significant parameters and updates, which complicates the definition of \emph{distortion}. In fact, in certain cases it has been shown that top $r$ sparsification outperforms non-sparse FL. Therefore, both top $r$ sparsification and random sparsification in private FL/FSL result in reduced communication costs, while the former suffers from increased storage cost and the latter suffers from decreased performance, i.e., accuracy of the model.

\subsubsection{Storage Constrained PRUW}

All of the previously mentioned PRUW schemes require databases, each with a storage capacity of at least the size of the training model. However, in practice, databases may not always have the space available to store an entire machine learning model. The given storage constraints of the databases can be either homogeneous or heterogeneous. The problem of private FSL with homogeneous storage constrained databases is studied in \cite{pruw}. The system model consists of $N$ non-colluding databases, each with a storage capacity of $\mu ML$, where $M$ and $L$ are the number of submodels and the size of a submodel and $\mu$ is a fraction in the range $\left[\frac{1}{N-3},1\right]$. Private FSL with heterogeneous storage constraints is studied in \cite{heterofsl}, where the storage capacity of database $n$ is given by $\mu_n ML$ for each $n$, where each $\mu_n\leq1$ is allowed to be arbitrary. The privacy, security and correctness constraints in both homogeneous and heterogeneous system models, are the same as in Section~\ref{pruw_fsl}.

There are two main methods to facilitate PRUW with storage constrained databases. 1) Storing only a fraction of each submodel in a given database, and replicating these fractions in only $r$, $r<N$ databases, i.e., \emph{divided storage}. 2) Using MDS codes to combine multiple model parameters and store them as a single symbol in each database, i.e., \emph{coded storage}. The schemes proposed in \cite{pruw} and \cite{heterofsl} for PRUW in FSL with storage constrained databases are based on the idea that combining both of the above methods, i.e., divided and coded storage, results in lower read-write costs, compared to only using either divided or coded storage. To this end, these schemes first find the optimum coding parameters and storage divisions that result in the minimum read-write costs for the given set of storage constraints, and use the basic PRUW scheme described in Section~\ref{pruw_fsl} to perform private FSL.

\section{Open Problems and Future Directions}

\begin{enumerate}
    \item Fundamental limits on performance metrics, i.e., converse results, have not yet been established for both PRUW and PSU based FSL. Investigating the fundamental limits is an interesting open problem, which might lead to capacity results of PRUW and PSU based FSL.
    
    \item A practical and interesting future direction related to both PRUW and PSU based FSL is to develop private read-write schemes that are robust against adversaries or/and eavesdroppers.
    
    \item All existing work on PRUW \cite{dropout, ourICC, pruw_journal, pruw, rps, model, heterofsl} consider a single client in the system model as shown in Fig.~\ref{pruwmodel}, i.e., all client-server communications are defined for a single client. A multi-client PRUW system model, where the private read-write process is defined from the perspective of a group of clients as opposed to defining it on an individual client with the same privacy, security and correctness requirements could be a promising future direction as it may result in a reduced communication cost, from combining queries/answers/updates of multiple clients. 
    
    \item In a setting where a given user has a wide variety of data to train multiple submodels in FSL at the same time, it would be interesting to see if there exists any method that decreases the total communication cost of privately reading/writing to multiple submodels simultaneously, compared to carrying out the PRUW process on each submodel individually.
    
    \item The rate-privacy-storage trade off in \cite{rps} only allows information to be leaked on the indices of the sparse updates in FL with sparsification, which does not have a significant impact on the communication rate. An interesting direction is to investigate the rate-privacy trade off in PRUW by incorporating differential privacy as in \cite{leaky,assymetric} to see the effect on the communication cost in private FSL/FL with sparsification when allowing a given amount of information to be leaked on the desired indices as well as the values of the selected updates/parameters.
    
    \item The concept of PRUW can be used in other applications such as private epidemiological data collection. This problem is introduced in \cite{epid} which contains three parties, namely, users, databases and the data collector. The problem setting contains both \emph{reading} and \emph{writing} components among different parties. The performance metric considered in \cite{epid} is the download cost at the data collector, i.e., the reading cost. An interesting future direction would be to consider the optimality of the writing cost, i.e., the upload cost of the users, which relates to the writing phase of PRUW.  
\end{enumerate}

\bibliographystyle{unsrt}
\bibliography{references}
\end{document}